\documentclass[11pt]{article}
\usepackage{amsmath}
\usepackage{amssymb}
\usepackage{graphicx}
\usepackage{lineno}
\usepackage{dcolumn}
\usepackage{bm}
\usepackage{setspace}

\usepackage{geometry}
\usepackage{indentfirst}
\usepackage{authblk}
\usepackage{hyperref}
\usepackage{booktabs}
\usepackage{cite}
\usepackage[bf,footnotesize]{caption}

\begin{document}
\newcommand\x{\textbf{x}}
\renewcommand\a{\textbf{a}}
\renewcommand\ss{\{0, 1\}^{N}}
\newcommand\C{\textbf{C}}
\newcommand\D{\textbf{D}}
\newcommand\A{\mathcal{A}}
\newcommand\G{\mathcal{G}}
\newcommand\T{\mathcal{T}}
\renewcommand\S{\mathcal{S}}

\renewcommand\c{\overline{c}}
\bibliographystyle{unsrt}
\title{Evolutionary dynamics on sequential temporal networks}

\author[ ]{Anzhi Sheng\textsuperscript{1}}
\author[ ]{Aming Li\textsuperscript{1,2,$*$}}
\author[ ]{Long Wang\textsuperscript{1,2,}\thanks{Corresponding authors: liaming@pku.edu.cn (A.L.), longwang@pku.edu.cn (L.W.)}}
\affil[1]{Center for Systems and Control, College of Engineering, \par{Peking University, Beijing 100871, China}}
\affil[2]{Center for Multi-Agent Research, Institute for Artificial Intelligence, \par{ Peking University, Beijing 100871, China}}
\renewcommand*{\Affilfont}{\small\it}
\renewcommand\Authands{ and }
\date{ }

\maketitle

\begin{abstract}
It is well-known that population structure is a catalyst for the evolution of cooperation since individuals can reciprocate with their neighbors through local interactions defined by network structures. Previous research typically relies on the assumption that population size is fixed and the structure is time-invariant, which is represented by a static network. However, real-world populations are often evolving with the successive growth of nodes and links in time, resulting in time-varying population structures. Here we model such growing networked populations by sequential temporal networks with an increasing number of nodes and edges and develop the theory of evolutionary dynamics on sequential temporal networks. We derive explicit conditions under which sequential temporal networks promote the evolution of cooperation relative to their static counterparts. In particular, even if natural selection disfavours cooperative behaviours on static networks, sequential temporal networks can surprisingly rescue cooperation. Furthermore, we demonstrate empirically that sequential temporal networks assembled from synthetic and empirical datasets present such promotion in the evolution of cooperation. Our results advance the study of evolutionary dynamics on temporal networks and open the avenue for investigating the evolution of prosocial and other behaviours.
\end{abstract}

\section{Introduction} 
Prosocial behaviours such as cooperation are ubiquitous ranging from microbial systems to human society \cite{Trivers64, Hofbauer1998, aexlrod1981}. 
Understanding the emergence and maintenance of cooperation has long been recognized as a significant problem because of its strong connection to the development of human societies \cite{keohane2016cooperation, block2020social}.
Evolutionary game theory is a powerful mathematical framework to study the evolution of cooperation.

Several mechanisms have been proposed in the literature \cite{nowak2006} to explore the emergence of cooperation, where population structure is one of the most important and widely discussed mechanisms \cite{nowak1992, szabo1998, hauert2004spatial, nowak2004emergence, santos2005, santos2008social, fu2008, tarnita2009, li2014cooperation, li2016evolutionary, allen2014, allen2017, allen2019, mcavoy2021, li2020, zhou2021}.  
Population structure is often modeled by a network, where nodes and edges represent individuals and mutual interactions, respectively. 
Individuals receive payoff through mutual interactions \cite{ohtsuki2006, su2019, nowak1992} defined by network structures.

An elementary assumption of previous studies is that evolutionary dynamics of cooperation occurs in fully evolved and fixed-size populations, meaning that the underlying network structure of populations is time-invariant.
Nevertheless, the evolution on networks is often coupled with the evolution of networks, most notably growth \cite{garcia2016hidden, hric2018stochastic}, in the real world.
Existing individuals interact with their neighbors through a network structure, while new individuals enter the population and connect to the existing individuals successively and form a new networked population.
The above pattern can be observed in plenty of complex systems -- such as information diffusion \cite{vazquez2006,vazquez2007,iribarren2009, davis2020phase}, where nodes enter a system sequentially when receiving information from spreaders, and the assembly of microbiome communities over time like gut microbiota aggregation within the gastrointestinal tract of infants\cite{stewart2018, rao2021, coyte2021}.
The growing process of networked populations has also been studied theoretically, using the master equation \cite{albert2000,dorogovtsev2000,goh2001universal} and branching growth \cite{zheng2021scaling}.
And one of the most famous models is the Barabási-Albert model \cite{barabasi1999}, where one adds a new node at each time step and links it to other nodes in the network with preferential attachment.
The evolution of cooperation in a single static network cannot capture the complexity and generality of evolutionary dynamics in growing populations with specific structures.
However, relevant research on this topic is still missing. 

Here we construct a sequential temporal network with an increasing number of nodes and edges and use it to describe a growing networked population with strategic evolution.
We study the evolution of cooperation on sequential temporal networks and quantify their ability to promote the evolution of cooperation and favour the fixation of cooperation. 
We provide mathematical conditions applicable to any sequential temporal network, under which sequential temporal networks have advantages in the evolution of cooperation over their corresponding static networks.
Analyses of four synthetic sequential temporal networks and four empirical sequential temporal networks demonstrate that sequential temporal networks are able to promote the evolution of cooperation. 
Furthermore, we propose a method to efficiently determine the superiority of sequential temporal over static networks in promoting cooperation.
Our results reveal the importance of population growth mechanisms for the evolution of cooperation in populations.

\section{Results}
\subsection{Model}
We model the interaction structure of a population with $N$ individuals by a network $\S$.
Each node is occupied by an individual and each edge describes a mutual interaction between two individuals. 
The static network $\S$ is specified by its adjacency matrix $W=(w_{ij})_{i,j=1}^{N}$, where $w_{ij}$ is the weight of edge $(i,j)$ representing the number of interactions per unit time. 

A growing population can be specified by a set of subnetworks (snapshots) in which the number of individuals is gradually increasing.
Such a set of networks is called a sequential temporal network.
In our model, the formation of a sequential temporal network $\T$ requires the input of a static network $\S$ with $N$ nodes and a vector set $\A=\{ \a^{(1)},...,\a^{(T)} \}$, i.e. $\T = (\S,\A)$.
The element $\a^{(t)}$ is a vector of length $N$, where $a_i^{(t)} = 1$ if node $i$ is activated at time $t$, otherwise $a_i^{(t)} = 0$.
The sequential temporal network has $T$ snapshots (i.e. $\T= \{\S^{(1)},...,\S^{(T)} \}$), where the activation of nodes in snapshot $\S^{(t)}$ is determined by $\a^{(t)}$. 
The evolution of the network stops when the structure is the same as $\S$ (i.e. $\S^{(T)}=\S$).
We show a more detailed construction of sequential temporal networks in Methods.

\begin{figure*}[p]
\centering
\includegraphics[width=1 \textwidth]{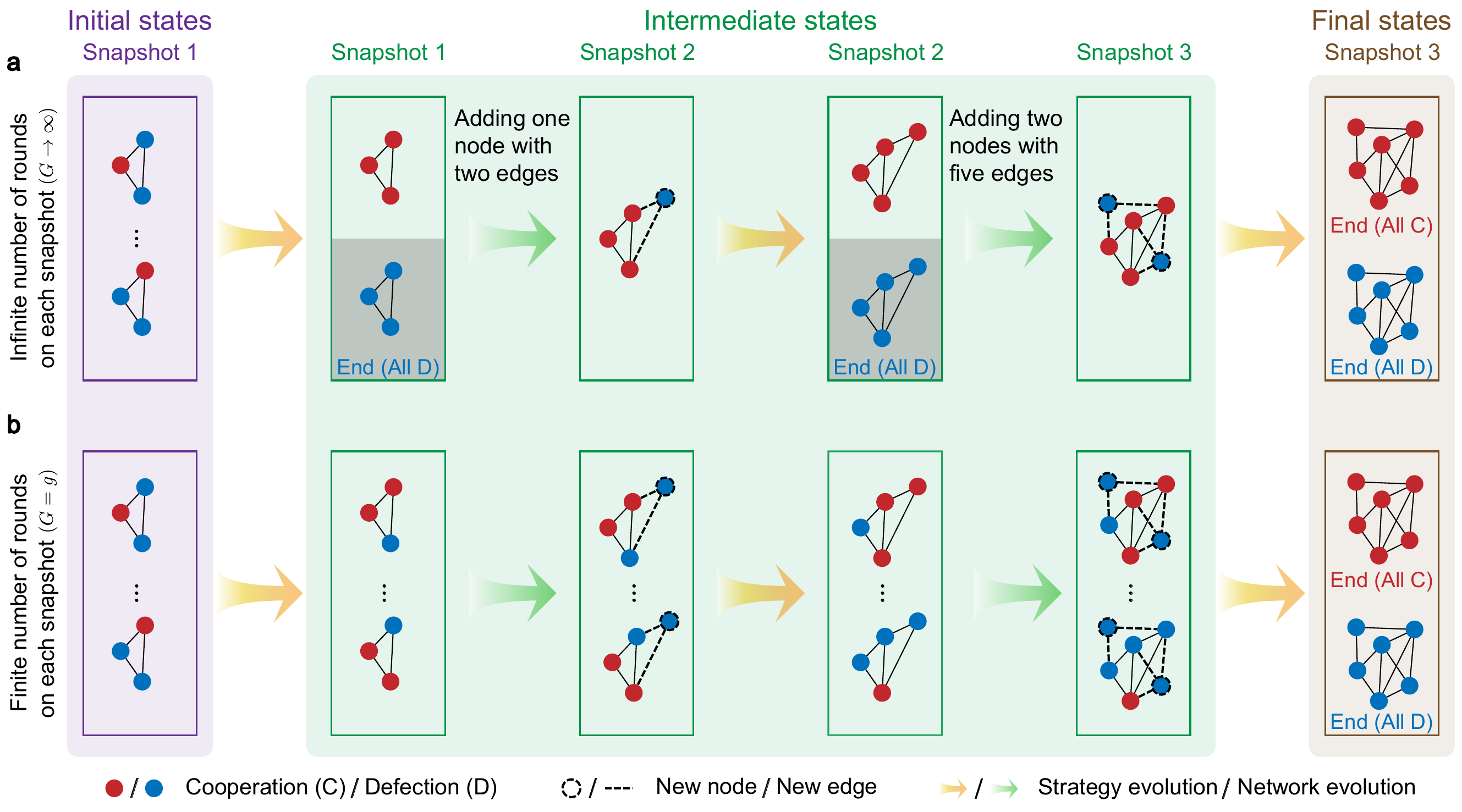}
\caption{\textbf{Illustration of two typical evolutionary processes on a sequential temporal network.} 
The sequential temporal network is formed by three snapshots.
The final snapshot is also the topology of its corresponding static network.
The evolutionary dynamics of cooperation and defection begins on the first snapshot, a three-node network.
The initial state is set randomly with a cooperator (C, red). 
After $G$ rounds of evolution, new nodes with strategy defection (D, blue) and edges (dashed lines) are added to the present network (intermediate states). 
The evolutionary process has two stable states (i.e. absorbing states): one in which all individuals become cooperators on the final snapshot (all C), and the other in which all individuals become defectors on any snapshot (all D, since newly added nodes are defectors).  
The evolutionary process ends when the population reaches one of the stable states. 
\textbf{a}, On each snapshot, the strategy of individuals is updated according to a given update rule until all individuals becomes cooperators or defectors, meaning that strategic evolution is sufficiently evolved. Then the network structure changes. 
\textbf{b}, On each snapshot except the last one, the evolution on the network proceeds $g$ rounds before switching to the next snapshot. 
On the last snapshot, the evolution continues until the whole network reaches an absorbing state. 
In particular, when $g\to \infty$, these two evolutionary processes are the same.
}
\end{figure*}

Individuals engage a two-player game in which both players can choose a strategy of cooperation (C) and defection (D) when interacting with each other.
Here we focus on the donation game \cite{ohtsuki2006}, in which cooperators pay a cost $c$ to donate $b$, and defectors pay no cost and provide no benefit.  
These outcomes can be represented by the following payoff matrix
\begin{spacing}{1.5}
	\centerline{$\bordermatrix{%
& \text{C} & \text{D} \cr
\text{C} & b-c & -c \cr
\text{D} & b & 0 \cr
}$.}
\end{spacing}
\noindent
Cooperators exhibit prosocial behaviours or spiteful behaviours \cite{forber2014evolution, mcavoy2020social} when $b>0,c>0$ or $b<0,c>0$.
In particular, when $b > c > 0$, this game is a Prisoners' Dilemma \cite{szabo1998}.

The state of a population with $N$ individuals is denoted by $\x = (x_1,...,x_N)^{\text{T}} \in \ss$, where $x_i=1$ ($x_i=0$) indicates that the strategy of individual $i$ is C (D). 
Each individual $i$ plays the game with each neighbor and receives an average payoff of $u_i(\x)=-c x_i + b \sum_{j=1}^{N} p_{ij}^{(1)} x_j$, where $p_{ij}^{(1)}=w_{ij} / \sum_{k=1}^{N} w_{ik}$ is a one-step random walk from $i$ to $j$. 
The fitness of individual $i$ is denoted by $F_i(\x) = 1 + \delta u_i(\x)$, where $\delta \ge  0$ is the intensity of selection \cite{nowak2004emergence}. 
The parameter $\delta=0$ corresponds to neutral drift and $\delta \ll 1$ corresponds to weak selection \cite{wu2010universality,wild2007different}.

The evolution of cooperation is driven by imitation. 
At each time step, a random individual $i$ is selected uniformly to update its strategy and copy the strategy of its neighbor $j$ with probability proportional to the edge-weighted fitness $w_{ji}F_j(\x)$.
This update rule illustrates that an individual tends to imitate the strategy of its successful neighbors.
We focus on this commonly used rule called death-birth updating \cite{ohtsuki2006, allen2017, su2019}, and we also analyse other update rules such as pairwise-comparison updating \cite{hauert2004spatial} and imitation updating \cite{ohtsuki2006} (see Supplementary Information section 2).
After a sufficient evolution, the state will reach $\textbf{C}=(1,...,1)^{\text{T}}$ (all C) or $\textbf{D}=(0,...,0)^{\text{T}}$ (all D), and these two states are called absorbing states.

We consider two different evolutionary processes on sequential temporal networks. 
Figure 1 illustrates the essence of these two processes. 
In the first evolutionary process, new node(s) with strategy defection will not enter the system until the state reaches absorbing states (Fig.~1a).
The evolution in each snapshot is sufficient, namely, the game is played in infinite rounds over each snapshot.
In the second process, the timescale of the evolutionary dynamics on each snapshot (except the last one) is controlled by the parameter $g$, which captures the number of rounds in each snapshot (Fig.~1b). 
In this case, the parameter $g$ determines the timescale difference between the evolution on the network and the evolution of the network.
When $g > 1$, the evolution on the network is faster than the evolution of the network.
When $g \to \infty$, the evolution on the network can be seen as reaching an equilibrium state in an instant based on the timescale of the evolution of the network, which is the same as the first evolutionary process.   
We first present the theoretical analysis and numerical simulations based on the first evolutionary process.

\subsection{General condition for the promotion of cooperation}
Considering the population eventually settles into $\textbf{C}$ or $\textbf{D}$, we quantify the ability of networks to facilitate the evolution of cooperation by the probability of reaching $\textbf{C}$, i.e. the fixation probability of cooperation \cite{ohtsuki2006, allen2017, mcavoy2020social, mcavoy2021}.
The fixation probability is a function of the initial configuration of cooperators and defectors on the network $\S$.
For a particular initial configuration $\boldsymbol{\xi}=(\xi_1,...,\xi_N)^{\text{T}} \in \ss$, the fixation probability of C is denoted by $\rho_{\S}({\boldsymbol{\xi}})$.
Another important initialization is called uniform initialization \cite{ohtsuki2006,su2019,allen2017} meaning that a single C is chosen uniformly at random in a population full of D, and the fixation probability of C, in this case, is denoted as $\rho_{\S}({\mu})$. 
In order to be consistent with the initialization of $\T$, the initialization of $\S$ is uniform.
For simplicity, we denote the fixation probability of a static network $\S$, $\rho_{\S}({\mu})$, by $\rho_{\S}$.
We indicate each variable under neutral drift and weak selection with a superscript $^\circ$ and $^*$, respectively.

We say that the sequential temporal network $\T = (\S,\A)$ promotes the evolution of cooperation relative to its static counterpart $\S$ if:
\begin{equation}
	\rho_{\T} > \rho_{\S}.
\end{equation}
Equation (1) shows that the probability of a single cooperator eventually taking over the population in $\T$ is higher than that in the corresponding static network $\S$.

We seek to derive the equivalent condition of equation (1).
The fixation probability of $\T=(\S,\A)$ can be formulated as 
\begin{equation}
	\rho_{\T} = \rho_{\S^{(1)}}({\mu})\prod_{t=2}^{T}\rho_{\S^{(t)}}({\boldsymbol{\xi}^{(t)}}),
\end{equation}
where $\rho_{\S^{(i)}}(\cdot)$ means the fixation probability of cooperation of snapshot $S^{(i)}$, $\boldsymbol{\xi}^{(t)}=\a^{(t-1)}$ ($2\le t\le T$) is a configuration, and $\a^{(t)}$ is an element of $\A$.
We first focus on unweighted sequential temporal networks with $T=2$, then equation (2) becomes $\rho_{\T}=\rho_{\S^{(1)}}({\mu})\rho_{\S^{(2)}}({\a^{(1)}})$, and $\rho_{\S}({\mu}) = \rho_{\S^{(2)}}({\mu})$.
As $\rho_{\T}^{*} > \rho_{\S}({\mu})^{*}$ can be deduced from $\rho_{\T}^{\circ} > \rho_{\S}({\mu})^{\circ}$, we first analyse the condition under neutral drift.
We assume that $\S^{(1)}$ ($\S^{(2)}$) has $m$ ($m+\Delta m$) nodes, the average connectivity of $\S^{(1)}$ ($\S^{(2)}$) is $k_1$ ($k_2$), and there are no interconnected edges among newly added nodes in $\S^{(2)}$.
Let $\Delta K$ denote the number of newly added edges in $\S^{(2)}$.
We obtain the identity $\Delta K=[(m+\Delta m)k_2-m k_1]/2$. 
Then equation (1) holds under neutral drift if and only if one of the following two conditions is satisfied:
\begin{equation} 
\begin{aligned}
 	&(\text{i})~\Delta m \ge m, \\
 	&(\text{ii})~\Delta m < m,\  \frac{\Delta K}{\Delta m}<\frac{m k_1}{m- \Delta m}.
\end{aligned}
 \end{equation}
When $\Delta m \ll m$, the second condition degenerates to 

\begin{equation*}
	(\text{ii})'~ \Delta m < m,\  \frac{\Delta K}{\Delta m}< k_1.
\end{equation*}

The condition (i) shows if the number of nodes added in the next snapshot ($\Delta m$) is not less than the original number ($m$), sequential temporal networks promote the evolution of cooperation.
The condition $(\text{ii})'$ shows when the average degree of newly added nodes ($\Delta K/ \Delta m$) is less than the average connectivity of the earlier snapshot ($k_1$), the cooperation is fostered by sequential temporal networks.
When the number of snapshots is greater than 3 (i.e. $T\ge 3$), a sufficient condition of $\rho_{\T}^{\circ} > \rho_{\S}^{\circ}$ is that each adjacent snapshots satisfy one of the conditions in equation (3).
Conversely, when none of adjacent snapshots satisfies equation (3), we have $\rho_{\T}^{\circ} \le \rho_{\S}^{\circ}$.
Figure 2 confirms the above conclusion.
All pairs of adjacent snapshots of the sequential temporal networks in Figs.~2a and 2b satisfy the condition (i) and (ii), respectively,
and those of the sequential temporal networks in Figs.~2c do not meet any of the conditions in equation (3).

\begin{figure*}[p]
\centering
\includegraphics[width=1 \textwidth]{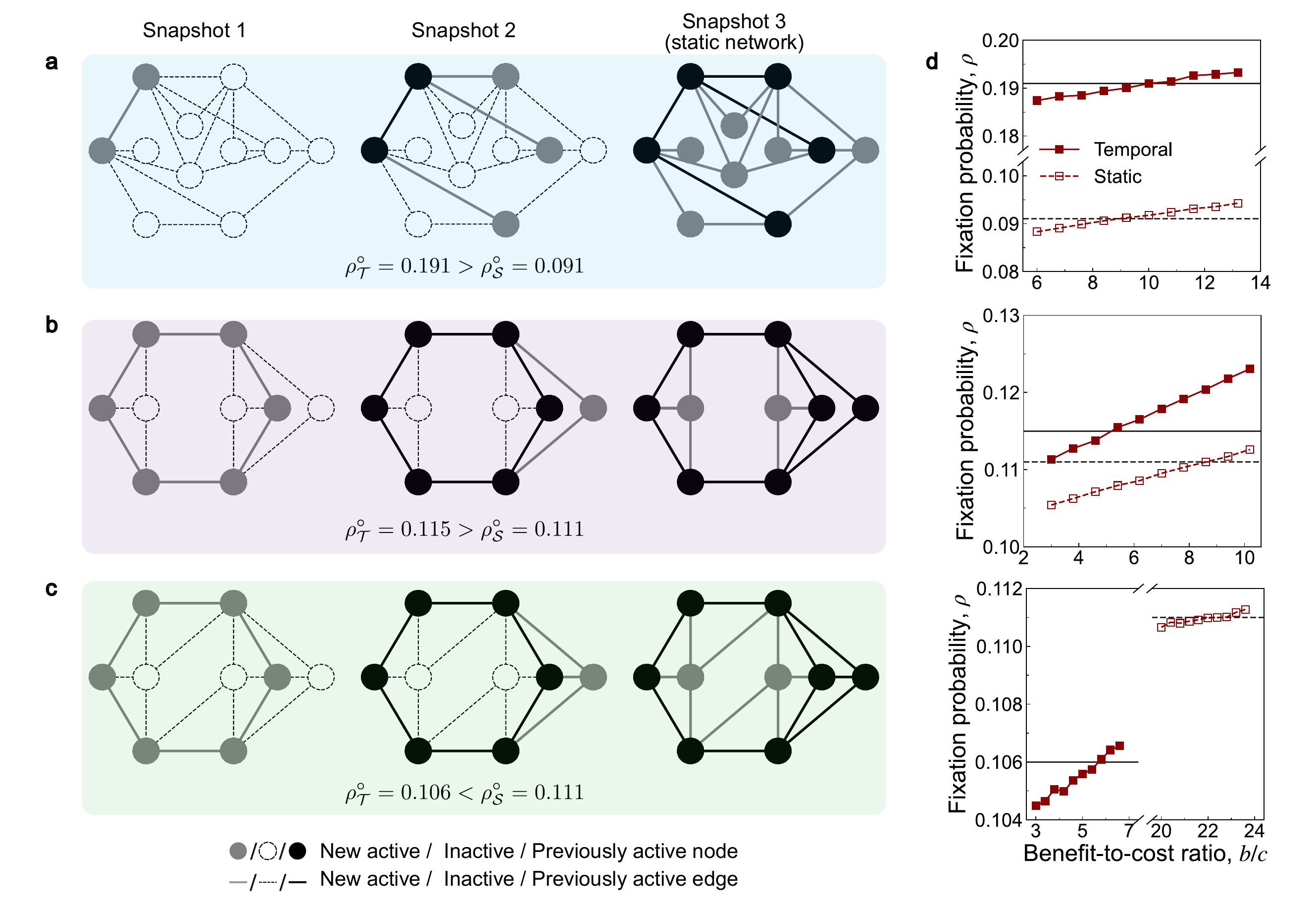}
\caption{\textbf{Fixation probability of cooperation of sequential temporal networks and their corresponding static networks.} 
We present three schematic examples to verify the theoretical results shown in equation (3).
Each sequential temporal network has three snapshots that form two pairs of adjacent snapshots.
The original number of nodes of the first and second pairs is denoted as $m_1$ and $m_2$, and the increasing number of nodes (edges) of the two pairs is denoted as $\Delta m_1$ ($\Delta K_1$) and $\Delta m_2$ ($\Delta K_2$), respectively.
\textbf{a}, The increase in the number of nodes of the two pairs of snapshots fulfills the condition (i) (i.e. $\Delta m_1 = 3 > m_1 = 2$ and $\Delta m_2 = 6 > m_2 = 5$). Then the fixation probability of sequential temporal network, $\rho_{\T}^{\circ} = 0.191$, is greater than that of its static counterpart, $\rho_{\S}^{\circ} = 0.091$.
\textbf{b}, The increase in the number of nodes and edges fulfills the condition (ii) (i.e. $\Delta m_1 = 1 < m_1 = 6$, $\Delta K_1 = 2 < 2.4$ and $\Delta m_2 = 2 < m_2 = 7$, $\Delta K_2 = 6 < 6.4$). As a result, the fixation probability of sequential temporal network, $\rho_{\T}^{\circ} = 0.115$, is greater than that of its static counterpart, $\rho_{\S}^{\circ} = 0.111$.
\textbf{c}, When each pair of adjacent snapshots does not satisfy the conditions (i) and (ii) (i.e. $\Delta m_1 = 1 < m_1 = 6$, $\Delta K_1 = 3 > 2.4$ and $\Delta m_2 = 2 < m_2 = 7$, $\Delta K_2 = 8 > 7.2$), the fixation probability of sequential temporal network, $\rho_{\T}^{\circ} = 0.106$, is smaller than that of its static counterpart, $\rho_{\S}^{\circ} = 0.111$.
\textbf{d}, The right-most column shows the result of the fixation probability of sequential temporal networks (squares) and their corresponding static networks (circles) under weak selection, which is obtained by $10^7$ replicate Monte Carlo simulations. 
The fixation probability of sequential temporal networks (static networks) under neutral drift is presented by horizontal solid (dashed) lines.
The inequality of fixation probabilities under neutral drift holds under weak selection.
Parameter values are $c=1$, $\delta=0.025$ for \textbf{a} and \textbf{b}, and $\delta=0.01$ for \textbf{c}.}
\end{figure*}

Applying equation (3), we find that the evolution of cooperation on sequential temporal networks is strongly correlated with the specific change in network topology over time.
When the number of nodes grows exponentially, the evolution of cooperation is promoted on sequential temporal networks.
When the growth rate of nodes ($\Delta m$) is slow, the increase of edges ($\Delta K$) needs to be upper bounded in order to foster the evolution of cooperation. 
Intuitively, since the newly added nodes are all defectors, it is essential to avoid the emergence of hubs from new nodes.
Therefore, the new nodes are not allowed to carry too many edges to enter the network.

\begin{figure}[t]
\includegraphics[width=1\textwidth]{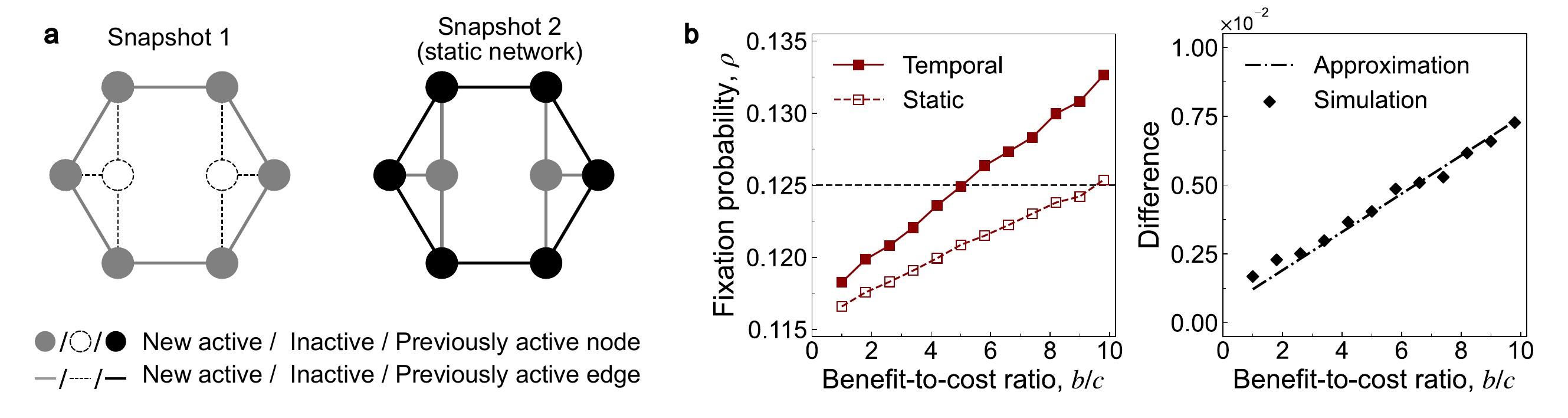}
\centering
\caption{\textbf{Fixation probability can be well estimated by the mean-field approximation.} 
\textbf{a}, We consider a sequential temporal network that has the same fixation probability of cooperation with its corresponding static network under neutral drift (i.e. $\rho_{\T}^{\circ}=\rho_{\S}^{\circ} = 0.125$, black dashed line).
\textbf{b}, The left panel presents the numerical simulation of fixation probabilities under weak selection, showing that $\rho_{\T}^{*}>\rho_{\S}^{*}$.
The difference between the sequential temporal network and the static network is plotted by black diomands in the right panel.
We approximate it by the difference between the left-hand side the right-hand side of equation (4) multiplying the intensity of selection $\delta$ (black dashed-dot line). As a result, the simulation is well estimated by the approximation. Parameter values are $c=1$, $\delta=0.025$.}
\end{figure}

When the fixation probability of a sequential temporal networks and its static counterpart is the same under neutral drift (i.e. $\rho_{\T}^{\circ} = \rho_{\S}^{\circ}$), we compare the first-order term of them under weak selection. 
We have derived the exact condition of equation (1) for any sequential temporal network (see Supplementary Information section 3.2 for more detailed derivations), but the complexity of verifying the condition is upper bounded by solving a linear system of size $O(T N^2)$, where $N$ is the size of static networks and $T$ is the length of the corresponding sequential temporal networks.
In fact, our main propose is to compare the magnitude of two sides of equation (1) rather than their specific difference.
Here we develop a mean-field approximation method \cite{fotouhi2019} to derive a computationally feasible condition to reduce the complexity (see Methods and Supplementary Information section 5 for details).
Applying the method, the approximate condition for $\rho_{\T}^{*} > \rho_{\S}^{*}$ with two snapshots is 
 \begin{equation}
\begin{aligned}
		\rho_{\S^{(1)}}(\mu)^{\circ} \cdot f^{\a^{(1)}}(\S^{(2)}) + \rho_{\S^{(2)}}(\a^{(1)})^{\circ}\cdot f^{\mu}(\S^{(1)}) > f^{\mu}(\S^{(2)}),
\end{aligned}
\end{equation}
where $f^{\mu}(\S^{(i)})$ and $f^{\boldsymbol{\xi}}(\S^{(i)})$ are the approximate perturbations on the fixation probability of $\S^{(i)}$ caused by individuals' payoffs with initialization $\mu$ and $\boldsymbol{\xi}$. The specific form of these notations can be found in Methods.
Figure 3 illustrates the validity of the mean-field approximation.
The fixation probability of the sequential temporal network is the same as that of the corresponding static network under neutral drift (i.e. $\rho_{\T}^{\circ} = \rho_{\S}^{\circ} = 0.125$),
but greater than that of the static network under weak selection (left panel of Fig.~4b).
Applying equation (4), we accurately predict the difference between $\rho_{\T}^*$ and $\rho_{\S}^*$ (right panel of Fig.~4b).

\subsection{General condition for the fixation of cooperation}

Selection is said to favour the fixation of cooperation on a network $\S$ when $\rho_{\S}^* > \rho_{\S}^{\circ}$ \cite{lieberman2005evolutionary, nowak2004emergence, ohtsuki2006}.
In the donation game, the above inequality is related to a critical value, $(b/c)^*$, which is known as the critical benefit-to-cost ratio \cite{allen2017, mcavoy2020social, su2019}. 
Positive critical ratios are lower bounds of the benefit-to-cost ratio $b/c$ to favour the fixation of cooperation, while negative critical ratios are upper bounds to favour the fixation of spite. 

Here we say that the sequential temporal network favours the fixation of cooperation by selection relative to its static counterpart if one of the following relations holds:

\begin{subequations}
	\begin{align}
		\left( \frac{b}{c} \right)^*_{\S} &> \left( \frac{b}{c} \right)^*_{\T} > 0, \tag{5a}\\
		\left( \frac{b}{c} \right)^*_{\S} &< 0 < \left( \frac{b}{c} \right)^*_{\T}. \tag{5b}
	\end{align}
\end{subequations}
 
Equation (5a) illustrates that sequential temporal networks decrease the required benefit-to-cost ratio to selectively favour the fixation of cooperation and equation (5b) shows that sequential temporal networks can favour cooperation even if the corresponding static networks favour spite.

The general equivalent condition of Eqs.~(5a) and (5b) can be obtained from the exact expression of fixation probability under weak selection (see Supplementary Information section 4 for more details).
In Fig.~4, we present three illustrative examples to show the advantage of the sequential temporal networks in the fixation of cooperation.
We consider $N=6$ individuals arranged in three different static networks. 
In Fig.~4a, the fixation of cooperation is favoured by selection only if $b/c$ exceeds $(b/c)^*_{\S} = 71.247$.
However, when the evolution occurs on the corresponding sequential temporal network, the critical value is reduced to $(b/c)^*_{\T} = 5.708$.
An even more interesting example is shown in Fig.~4b. 
The critical benefit-to-cost ratio is infinite, $(b/c)^*_{\S}=+\infty$, meaning that cooperation is never favoured by selection.
Nevertheless, the critical value of the corresponding sequential temporal network can decrease to a finite value even if the critical value of the first snapshot is infinite. 

\begin{figure}[t]
\centering
\includegraphics[width=1 \textwidth]{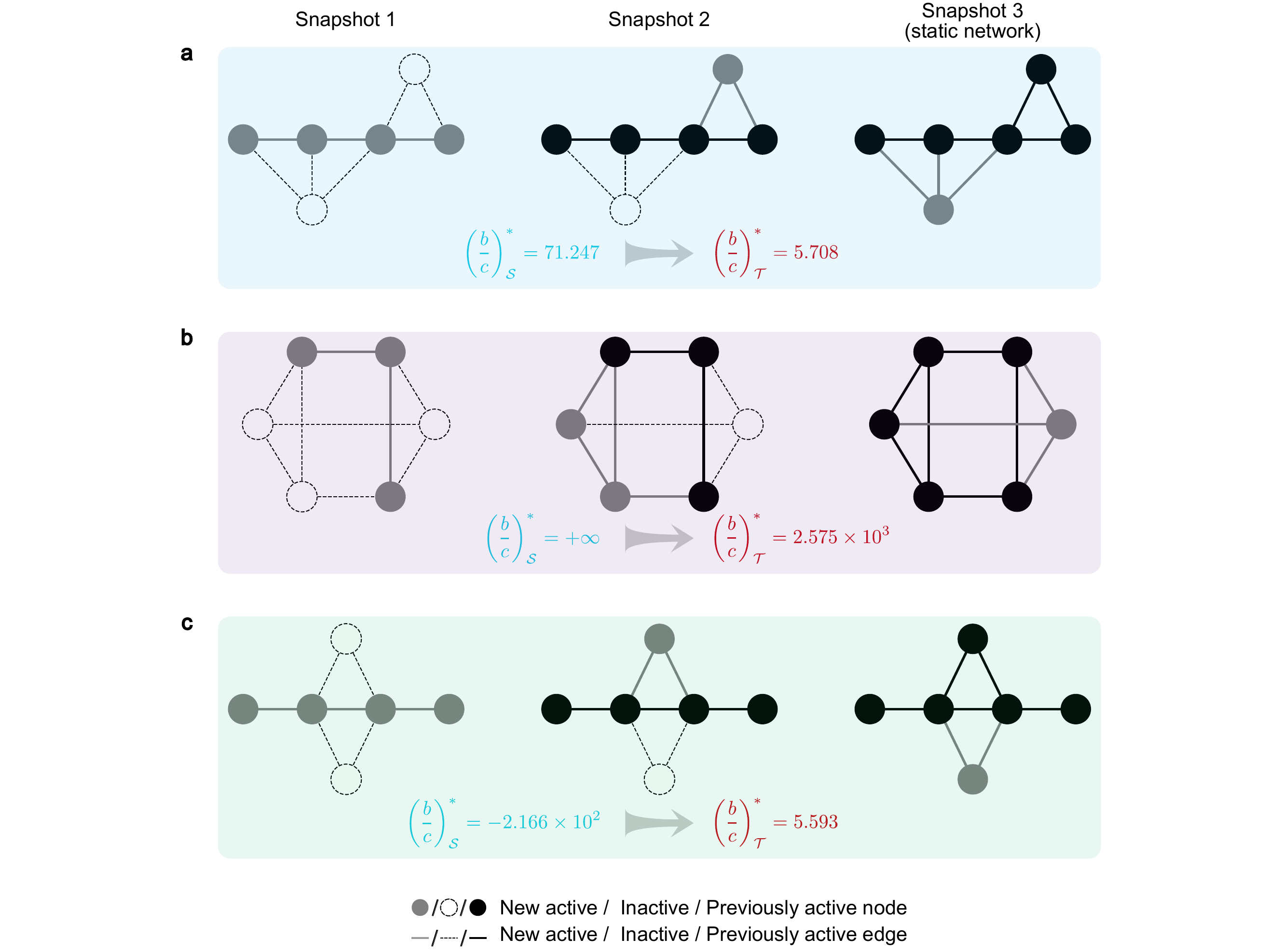}
\caption{\textbf{Sequential temporal networks can favour the fixation of cooperation.} We present three schematic examples.
\textbf{a}, The critical benefit-to-cost ratio of the static network is positive but higher than that of the sequential temporal network, i.e. $(b/c)^*_{\S} = 71.247 > (b/c)^*_{\T} = 5.708$, so that cooperators can provide less donation for being favoured by selection in the sequential temporal network. 
\textbf{b}, The critical benefit-to-cost ratio of the static network is infinite, i.e. $(b/c)^*_{\S} = + \infty$.
It follows that cooperation is never favoured by selection. However, selection can favour the fixation of cooperation in the sequential temporal network, provided the benefit-to-cost ratio, $b/c$, is larger than the critical value, $(b/c)^*_{\T} = 2.575 \times 10^3$.
\textbf{c}, The critical benefit-to-cost ratio of the static network is negative, i.e. $(b/c)^*_{\S} = -2.166 \times 10^2 < 0$. As a result, selection favour the fixation of spiteful behaviours, meaning that individuals pay a cost, $c>0$, to decrease its neighbors' payoff. Nevertheless, when the evolutionary dynamics is on the sequential temporal network, selection can favour the fixation of cooperation because of $(b/c)^*_{\T} = 5.593 > 0$.}
\end{figure}

The two examples above all fulfill equation (5a). Figure 4c shows an example satisfying equation (5b).
The critical benefit-to-cost ratio of the static network is negative, $(b/c)^*_{\S}<0$, which means that selection favours spiteful behaviours.
But if we consider the evolutionary dynamics on the sequential temporal network, the critical value becomes positive, indicating that selection favours cooperation. 

Similar to the fixation probability, the computational consumption of the critical benefit-to-cost ratio is high when faced with large static networks or long sequential temporal networks.
We also use the mean-field approximation mentioned above to obtain the critical benefit-to-cost ratio (see Methods).

A natural question is whether there exists a sequential temporal network that both promotes the evolution of cooperation and favours the fixation of cooperation by selection.
The sequential temporal network presented in Fig.~2b is a perfect example to answer the question.
The fixation probability of the sequential temporal network is greater than that of its static counterpart under neutral drift, $\rho_{\text{C}}(\T)^{\circ} = 0.115 > \rho_{\text{C}}(\S;\mu)^{\circ} = 0.111$, and the critical benefit-to-cost ratio of the the sequential temporal network is smaller than that of its static counterpart, $(b/c)^*_{\T} = 4.555 < (b/c)^*_{\S}=8.388$.
When the structure of sequential temporal networks is more complicated, we can still find examples that are superior in both of these two metrics (see Fig.~6b and Supplementary Fig.~3).

\begin{figure}[t]
\includegraphics[width=1 \textwidth]{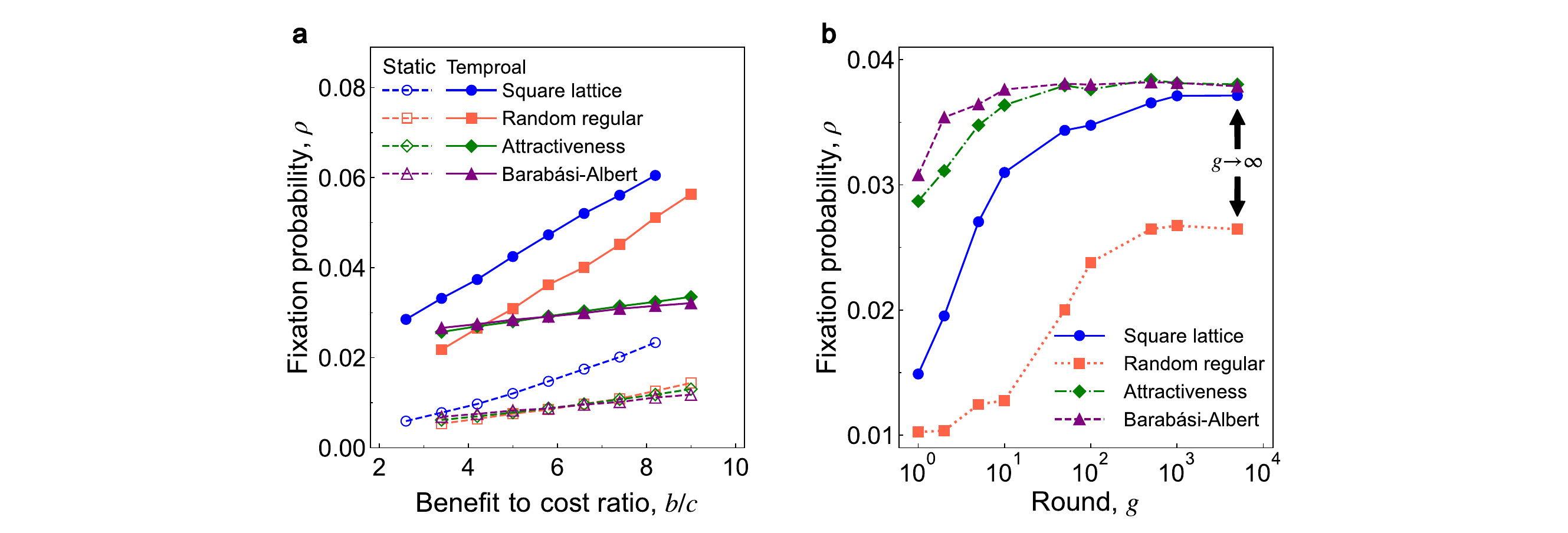}
\centering
\caption{\textbf{Evolution of cooperation on synthetic networks.} Networks are square lattices, random regular graphs with average connectivity $k$, scale-free networks with initial attractiveness $a$ and linking number $m$, and scale-free networks generated by the Barabási-Albert model with linking number $m$.
We obtain the numerical simulation of fixation probabilities by averaging over $10^6$ independent Monte Carlo simulations.
\textbf{a}, Based on the first evolutionary process, the fixation probability of the sequential temporal networks (solid lines) is greater than that of their corresponding static counterparts (dashed lines) under weak selection.
\textbf{b}, Based on the second evolutionary process, the fixation probability of the sequential temporal networks is monotonically increasing with respect to $g$ under neutral drift. Furthermore, we notice that the fixation probabilities are greater than $0.01$ when $g \ge 1$, meaning that these sequential temporal networks promote the evolution of cooperation even if the evolution on each snapshot (except the last one) is not sufficient. 
All static networks have size $N=100$ and other parameter values are $c=1$, $\delta=0.025$, $k=6$, $a=50$ and $m=3$.}
\end{figure}
\subsection{Synthetic and empirical temporal networks}
The sequential temporal networks discussed above are relatively short and small, but they nonetheless present a striking effect on the evolution of cooperation.
Here we study the evolutionary dynamics on larger networks, of size $N=100$ and their corresponding sequential temporal networks are longer, of length $T \ge 95$.
We selected four classic networks, which are square lattices with periodic boundaries (SL) \cite{nowak1992}, random regular graphs (RR) \cite{steger1999generating}, scale-free networks with initial attractiveness (IA) \cite{dorogovtsev2000} and scale-free networks generated by the Barabási-Albert model (BA) \cite{barabasi1999}, and generated the corresponding sequential temporal networks (see Supplementary Information section 7.1 for detailed constructions).
The former two networks are homogeneous but have very different local structures (such as the clustering coefficient), and the latter two networks are heterogeneous with different scaling laws.
In Fig.~5a, we show the fixation probability of cooperation of these four static networks and corresponding sequential temporal networks under weak selection.
The fixation probability of the sequential temporal networks is greater than that of the static networks, meaning that the sequential temporal networks promote the evolution of cooperation.
Applying the mean-field approximation, we obtain the critical benefit-to-cost ratio of the static networks and sequential temporal networks (see Supplementary Fig.~3 and Table 1).
All critical values are larger than $0$, meaning that the fixation probabilities are all monotonically increasing with respect to the benefit-to-cost ratio $b/c$.
It is worth noting that equation (5a) holds for the random regular graph, which shows that the sequential temporal version of the random regular graph both promotes the evolution of cooperation and favours the fixation of cooperation.

We turn to study the second evolutionary process. 
We investigate the relationship between the fixation probability, $\rho_{\T}(g)$, and the parameter $g$ on the four sequential temporal networks under neutral drift.
Figure 5b shows how the cooperation evolves when the number of rounds $g$ over each snapshot changes. 
The fixation probability of these four sequential temporal networks increases monotonically with respect to $g$ and converges to the value on the first evolutionary process as $g \to \infty$.  
Similarly, the monotonicity is also determined by the structure of sequential temporal networks (see Supplementary Information section 6 for detailed derivations). 

We notice that the fixation probability, $\rho_{\T}(1)^{\circ}$, of SL networks, RR graphs, IA networks and BA networks equals $1.489\times 10^{-2}$, $1.026\times 10^{-2}$, $2.869\times 10^{-2}$ and $3.078\times 10^{-2}$, respectively, which are all higher than the fixation probability of their static counterparts ($\rho_{\S}^{\circ} = 0.01$).
Therefore, to foster the evolution of cooperation, the dynamics of strategies does not need to be fully evolved on every snapshot.
Intuitively, the parameter $g$ affects the expected time of reaching $\textbf{C}$ (i.e. conditional absorbing time of $\textbf{C}$).
This raises the question of whether there exists $g$ to balance the promotion of cooperation and the absorbing time of reaching $\textbf{C}$.
In fact, we find such a tradeoff on these four sequential temporal networks when we set $g=10$ (see Supplementary Fig.~4). 
In this way, the fixation probability is higher and the conditional absorbing time of $\textbf{C}$ is lower than the static networks.

\begin{figure}[p]
\includegraphics[width=1 \textwidth]{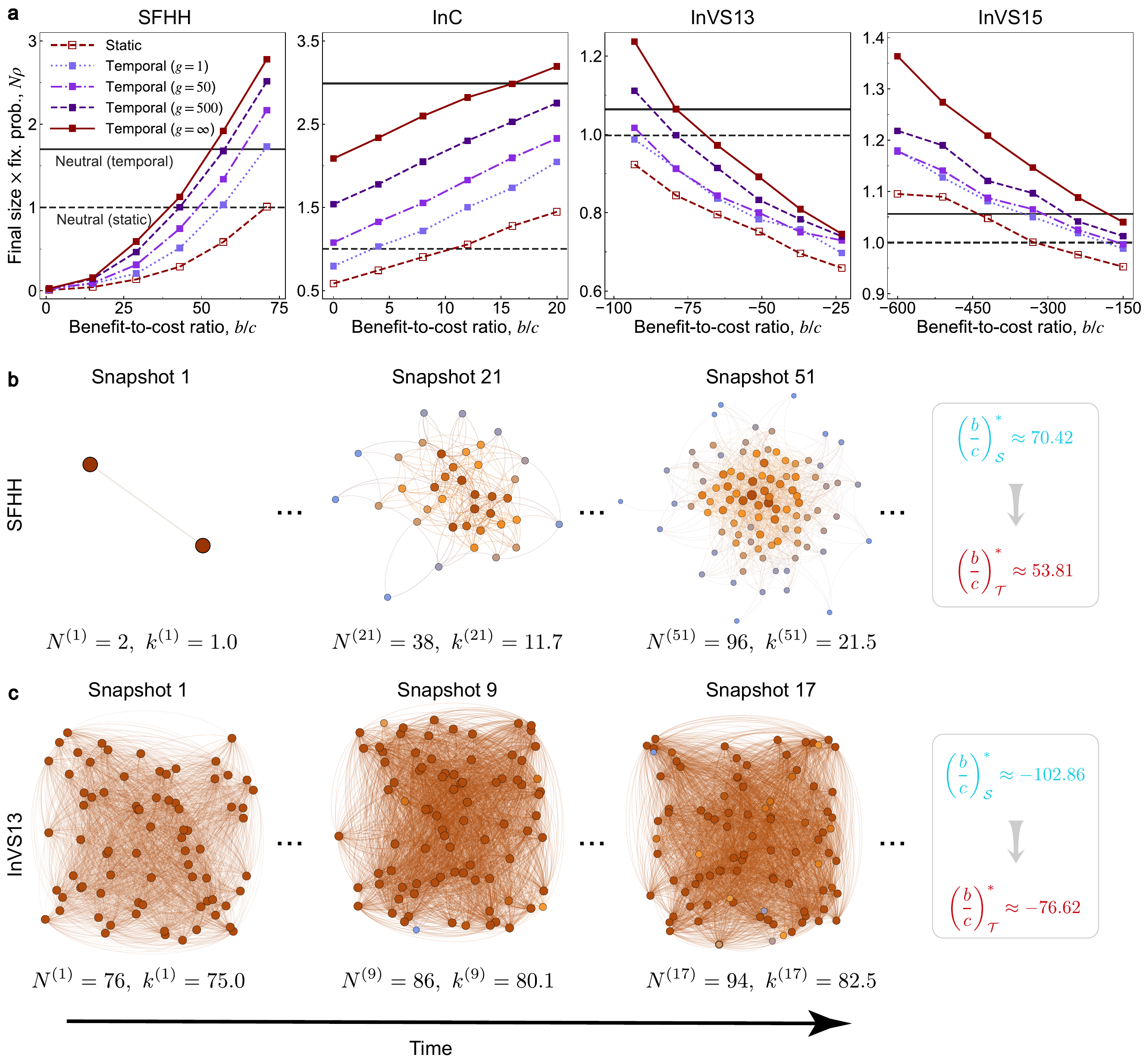}
\centering
\caption{\textbf{Evolution of cooperation in four empirical datasets.}
The datasets are collected from different social contexts: a scientific conference in Nice, France (SFHH) \cite{genois2018}, the Science Gallery in Dublin, Ireland (InC) \cite{isella2011}, a workplace with data collected in two different years in an office building in France (InVS13, InVS15) \cite{genois2015data}.
\textbf{a}, We consider the evolution of cooperation based on both the first and second evolutionary processes, where individuals play the donation game. The fixation probability of the empirical sequential temporal networks (squares) is greater than that of the corresponding static networks (circles) for any $g \ge 1$ under both neutral drift and weak selection.
\textbf{b},\textbf{c}, Schematic representation of snapshots at different moments, $N^{(t)}$ and $k^{(t)}$ indicate the size and average degree of snapshot $\S^{(t)}$, respectively.  \textbf{b}, Except for promoting the evolution of cooperation, the sequential temporal network of the SFHH dataset also favours the fixation of cooperation, i.e. $(b/c)^*_{\T} \approx 53.81 < (b/c)^*_{\S} \approx 70.42$.
\textbf{c}, The approximate critical benefit-to-cost ratio of the sequential temporal network and the static network of the InVS13 dataset is negative, i.e. $(b/c)^*_{\T} \approx -76.62 < 0$,  $(b/c)^*_{\S} \approx -102.86  < 0$. And the relation $|(b/c)^*_{\T}| < |(b/c)^*_{\S}|$ holds. As a result, the sequential temporal network favours the fixation of spite.   
Parameter values are $c=1$ in all datasets, $\delta=0.025$ in SFHH, and $\delta=0.01$ in the rest of datasets.}
\end{figure}

Finally, we investigate the evolution of cooperation on empirical networks from SocioPatterns \cite{web}.
We construct four empirical static networks and corresponding sequential temporal networks assembled from four empirical datasets that collect social interactions from different social contexts \cite{genois2015data, genois2018, isella2011} (see Supplementary Information section 7.2 for detailed constructions) and analyse strategic evolution on these networks.
The detailed information of these networks is listed in Supplementary Table 2.

Figure 6a shows the fixation probability of cooperation on the static networks and the sequential temporal networks under weak selection.
In these four datasets, the sequential temporal networks facilitate the evolution of cooperation even if the evolution on each snapshot is not sufficient.
We also observe that the fixation probability of these sequential temporal networks is monotonically increasing with respect to $g$. 

We notice that the monotonicity of the fixation probability with respect to the benefit-to-cost ratio, $b/c$, is different in the first two networks (SFHH and InC) than in the last two networks (InVS13 and InVS15), which is controlled by the sign of the critical benefit-to-cost ratio.
Furthermore, we use the mean-field approximation to estimate the critical benefit-to-cost ratio of networks.
The critical value of the sequential temporal network ($(b/c)^*_{\T} \approx 53.81$) is lower than that of the corresponding static network ($(b/c)^*_{\S} \approx 70.42$) in SFHH dataset (Fig.~6b), meaning that the sequential temporal network of SFHH favours the fixation of cooperation.
For InVS13 (Fig.~6c) and InVS15 datasets, the critical benefit-to-cost ratio of temporal and static networks is all negative, but the absolute value of the sequential temporal networks is relatively small, which indicates that the sequential temporal networks favour the fixation of spite relative to its static counterparts. 

\section{Discussion}

In this work, we study the evolution of cooperation in growing networked populations modeled by sequential temporal networks where new nodes and edges successively enter the networks.
Each snapshot of sequential temporal networks records the specific topology of the corresponding static networks at each time step.
We enumerate two typical evolutionary processes on sequential temporal networks to describe the coupling between the evolution on networks and the evolution of networks.

Our results offer a new insight to understand the evolution of prosocial behaviours in networked populations. 
By analyzing synthetic and empirical datasets, we show the evident advantages of sequential temporal networks in promoting the evolution of cooperation and reducing the conditional absorbing time.
A recent study demonstrates that selection will not favour the fixation of cooperation on roughly one-third of static networks \cite{allen2017}. 
Interestingly, we find that the corresponding sequential temporal networks can rescue cooperation in these static networks.
Specifically, we present several examples to show that sequential temporal networks can support the fixation of cooperation even though cooperation is never favoured or spite is favoured by selection on traditional static networks. 

We demonstrate that the advantages of sequential temporal networks are caused by the systematical growth of nodes and edges during the network evolution, and provide a general rule of population growth to facilitate the evolution of cooperation.
Similarly, several important prior studies have also considered some specific rules driven by evolutionary dynamics for population growth and showed that the evolution of cooperation is significantly influenced by the population growth \cite{poncela2008complex, poncela2009evolutionary}.
These growth rules can be generally viewed as special cases in our framework, since our rules only specify the relationship between the number of newly added nodes and edges, independent of how they are connected.

Our work also provides a method to efficiently calculate the evolutionary results of static and sequential temporal networks (Supplementary Figs.~4 and 6). 
Compared to traditional solutions, this method has a significant advantage in terms of time consumption, especially when the network size is large.
This method can also be applied for different updating rules (see Supplementary Note 5).

We have focused on pairwise interactions in networks, where individuals engage a two-player game.
A natural extension is to consider higher-order interactions \cite{battiston2020networks, alvarez2021evolutionary, battiston2021physics, lambiotte2019networks} or group interactions \cite{mcavoy2020social, santos2008social, perc2013evolutionary} in networks, since cooperation may unfold in groups. 
In this way, several newly added nodes with specific structures will enter networked systems as a whole, and individuals may simultaneously engage in both two-player and multi-player games with different opponents. 
Overall, after uncovering many surprising properties of the evolution of cooperation on sequential temporal networks, we believe that our findings deepen the understanding of the importance of network evolution for the fate of cooperators.

\section{Methods}

\subsection{Sequential temporal network construction}
The construction of a sequential temporal network is based on a static network $\S$ with an adjacency matrix $W=(w_{ij})_{i,j=1}^{N}$ and a set of activation vectors, $\A=\{ \a^{(1)},...,\a^{(T)} \}$.
In this case, the sequential temporal network is denoted as $\T= \{\S^{(1)},...,\S^{(T)} \}$,  where the number of nodes in the snapshot $\S^{(t)}$ is $\sum_{i=1}^{N} a^{(t)}_i$, and the adjacency matrix of snapshot $\S^{(t)}$ is the connected part of the matrix $W^{(t)} = \text{diag}(\a^{(t)})\cdot W \cdot \text{diag}(\a^{(t)})$.
We define a partial ordering $\preccurlyeq$ on $\mathbb{R}^N$. The relation $x=(x_i)_{i=1}^{N} \preccurlyeq y=(y_i)_{i=1}^{N}$ holds when $x_i \le y_i$ for all $1 \le i \le N$.
To satisfy the definition of a sequential temporal network, the relation $\a^{(t_1)} \preccurlyeq \a^{(t_2)}$ holds for all $1 \le t_1\le t_2 \le T$.
Furthermore, we set $\a^{(T)} = (1,...,1)^{\text{T}}$ (i.e. $\S^{(T)}=\S$), which means that the evolution of populations stops when the population structure is the same as the pre-given static network $\S$.
\subsection{Mean-field approximation}
Here we briefly summarize the mean-field approximation of the fixation probability and the critical benefit-to-cost ratio of static networks and sequential temporal networks under DB updating.
The detailed derivations of the approximation and the result of other update rules can be found in Supplementary Information section 5.

Each snapshot with $N$ nodes is described by an undirected graph $\S$ with weights $w_{ij}$ ($w_{ij} = w_{ji}$ for all $i,j$) and no self-loops ($w_{ii}=0$ for all $i$).
The weighted degree of node $i$ is $w_i = \sum_{j=1}^{N} w_{ij}$, and the probability of node $i$ taking $n$ steps to node $j$ is denoted as $p_{ij}^{(n)}$.
The reproductive value of node $i$ is $\pi_i = w_i/\sum_{k=1}^{N} w_k$ \cite{mcavoy2021}, which is the invariant distribution of randoms walks on the graph
For any vector $\textbf{y}=(y_1,...,y_N)^{\text{T}}$ on $\S$, we define the RV-weighted value $\widehat{y}:=\sum_{i=1}^{N}\pi_i y_i$.

\subsubsection{Fixation probability} 
For a particular initial configuration $\boldsymbol{\xi}=(\xi_1,...,\xi_N)^{\text{T}}$, let $t_{ij} = \frac{N}{2}(\widehat{\xi} - \xi_i \xi_j)$ and $B_0 = \sum_{i,j = 1}^{N} \pi_{i} \pi_j t_{ij}$. 
The mean-field approximation of the fixation probabilities $\rho_{\S}({\boldsymbol{\xi}})^*$ and $\rho_{\S}(\mu)^*$ is given as

\begin{equation}
	\begin{aligned}
		\rho_{\S}({\boldsymbol{\xi}})^* &\approx \widehat{\xi} + \frac{\delta}{N} \left( -c \cdot \mathcal{C}^{\boldsymbol{\xi}}(\S) +b \cdot \mathcal{B}^{\boldsymbol{\xi}}(\S)  \right) + O(\delta^2) \\
		&=  \widehat{\xi} + \frac{\delta}{N} \left( -c\gamma_{(2)}^{\boldsymbol{\xi}} + b(\gamma_{(3)}^{\boldsymbol{\xi}} - \gamma_{(1)}^{\boldsymbol{\xi}})  \right) + O(\delta^2) \\
		&= \widehat{\xi} + \frac{\delta}{N} \left( -c \left(\frac{B_0N \mu_1^2}{\mu_2} - \frac{1}{N\mu_1}(C_0 + C_1) \right) \right. \\
		&+b \left. \left( \frac{B_0 \Lambda \mu_1}{\mu_2} - \frac{1}{N\mu_1}(C_1+C_2)\right) \right)+ O(\delta^2) \\
		\rho_{\S}(\mu)^* &\approx \frac{1}{N} +  \frac{\delta}{N}\left( -c \cdot \mathcal{C}^{\mu}(\S) +b \cdot \mathcal{B}^{\mu}(\S) \right) + O(\delta^2) \\
		&= \frac{1}{N} +  \frac{\delta}{N}\left( -c \gamma_{(2)}^{\mu} + b(\gamma_{(3)}^{\mu} - \gamma_{(1)}^{\mu}) \right) + O(\delta^2)\\
		&= \frac{1}{N} + \frac{\delta}{N} \left(-c\left(\frac{N \mu_1^2}{2\mu_2} - 1\right)+b\left(\frac{\Lambda \mu_1}{2\mu_2} - 1\right)\right) + O(\delta^2),
	\end{aligned}
\end{equation}
where 
\begin{equation}
	\begin{aligned}
	\gamma_{(1)}^{\boldsymbol{\xi}} &= \frac{B_0N \mu_1^2}{\mu_2} - \frac{1}{N\mu_1} \sum_{i=1}^{N} w_i t_{ii}, \\
	\gamma_{(2)}^{\boldsymbol{\xi}} &= \frac{B_0N \mu_1^2}{\mu_2} - \frac{1}{N\mu_1}\left( \sum_{i,j=1}^N w_i (p_{ij}^{(0)}+ p_{ij}^{(1)}) t_{ij} \right), \\
	\gamma_{(3)}^{\boldsymbol{\xi}} &= \frac{B_0N \mu_1^2}{\mu_2} + \frac{B_0 \mu_1 \sum_{i=1}^N w_i p_{ii}^{(2)}}{\mu_2} - \frac{1}{N\mu_1} \left( \sum_{i,j=1}^N w_i (p_{ij}^{(0)}+ p_{ij}^{(1)} + p_{ij}^{(2)}) t_{ij} \right), \\
	\gamma_{(1)}^{\mu} &= \frac{N \mu_1^2}{2\mu_2} - \frac{1}{2}, \\
	\gamma_{(2)}^{\mu} &= \frac{N \mu_1^2}{2\mu_2} - 1, \\
	\gamma_{(3)}^{\mu} &= \frac{N \mu_1^2}{2\mu_2} + \frac{\sum_{i=1}^N w_i p_{ii}^{(2)} \mu_1}{2\mu_2} - \frac{3}{2}, \\
	\end{aligned}
\end{equation}
$\mu_1 = \sum_{i=1}^N w_i / N$ and $\mu_2 = \sum_{i=1}^N w_i^2 / N$ are the first and second moments of the network weighted degree distribution, $C_k = \sum_{i,j=1}^{N} w_i p_{ij}^{(k)} t_{ij}$ for $k=0,1,2$, and $\Lambda = \sum_{i=1}^{N}w_i p_{ii}^{(2)}$.
All these notations are allowed to be calculated without solving linear systems (see Supplementary Information section 2).
Applying equation (6), the perturbation on fixation probabilities is approximated as

\begin{equation}
	\begin{aligned}
		f^{\boldsymbol{\xi}}(\S) &= \frac{1}{\mathcal{N}(\S)} \left(-c \cdot \mathcal{C}^{\boldsymbol{\xi}}(\S) +b \cdot \mathcal{B}^{\boldsymbol{\xi}}(\S) \right), \quad
		f^{\mu}(\S) &= \frac{1}{\mathcal{N}(\S)} \left(-c \cdot \mathcal{C}^{\mu}(\S) +b \cdot \mathcal{B}^{\mu}(\S) \right),
	\end{aligned}
\end{equation}
where the notation $\mathcal{N}(\S)$ indicates the network size of $\S$.
Furthermore, we can also obtain the approximate fixation probability of sequential temporal networks.

\subsubsection{Critical benefit-to-cost ratio}
The mean-field approximation of the approximate critical benefit-to-cost ratio of a static network with an arbitrary initial configuration $\boldsymbol{\xi}$ and uniform initialization is given by  
\begin{equation}
	\left( \frac{b}{c} \right)^*_{\boldsymbol{\xi}} \approx \frac{B_0 N^2 \mu_1^2 - \mu_2(C_0 + C_1)}{B_0 N \Lambda \mu_1 - \mu_2(C_1 + C_2)},
\end{equation}
and
\begin{equation}
	\left( \frac{b}{c} \right)^*_{\mu} \approx \frac{N \mu_1^2 - 2\mu_2}{\Lambda \mu_1 - 2\mu_2},
\end{equation}
respectively, where the notations are the same as those in equation (6).

For a sequential temporal network $\T = (\S, \A)$, let  $A_1 = \rho_{\S^{(1)}}(\mu)^{\circ}$, $A_i = \rho_{\S^{(i)}}(\a^{(i-1)})^{\circ}$ ($i = 2,...,T$) and $\widetilde{A}_{i} = \prod_{j\neq i} A_j$. 
The approximate critical value is given as 
\begin{equation}
	\left( \frac{b}{c} \right)^*_{\T} = \frac{\widetilde{A}_1 \left(\frac{\gamma^{\mu}_{(2)}}{N} \right)\bigg|_{\S^{(1)}} +\left(\sum_{i=2}^{T} \widetilde{A}_i \left(\frac{\gamma^{\a^{(i-1)}}_{(2)}}{N} \right)\bigg|_{\S^{(i)}} \right)}{\widetilde{A}_1 \left(\frac{\gamma^{\mu}_{(3)}-\gamma^{\mu}_{(1)}}{N} \right)\bigg|_{\S^{(1)}} + \left(\sum_{i=2}^{T} \widetilde{A}_i \left(\frac{\gamma^{\a^{(i-1)}}_{(3)}-\gamma^{\a^{(i-1)}}_{(1)}}{N} \right)\bigg|_{\S^{(i)}} \right)},
\end{equation}
where the notation $\cdot|_{\S}$ indicates that the value is taken under $\S$, and the notations $\gamma^{\mu}_{(i)}, \gamma^{\a^{(j)}}_{(i)}$ ($i=1,2,3$ and $j=1,...,T-1$) are mentioned in equation (7).

\end{document}